\documentstyle[12pt]{article}

\makeatletter
\def\eqnarray{\stepcounter{equation}\let\@currentlabel=\theequation
\global\@eqnswtrue
\global\@eqcnt\z@\tabskip\@centering\let\\=\@eqncr
$$\halign to \displaywidth\bgroup\@eqnsel\hskip\@centering
  $\displaystyle\tabskip\z@{##}$&\global\@eqcnt\@ne 
  \hfil$\displaystyle{\hbox{}##\hbox{}}$\hfil
  &\global\@eqcnt\tw@ $\displaystyle\tabskip\z@
  {##}$\hfil\tabskip\@centering&\llap{##}\tabskip\z@\cr}
\def\@sect#1#2#3#4#5#6[#7]#8{\ifnum #2>\c@secnumdepth
    \def\@svsec{}\else
    \refstepcounter{#1}\edef\@svsec{\csname the#1\endcsname.\hskip 1em }\fi
    \@tempskipa #5\relax
    \ifdim \@tempskipa>\z@
    \begingroup #6\relax
    \@hangfrom{\hskip #3\relax\@svsec}{\interlinepenalty \@M #8\par}
    \endgroup
    \csname #1mark\endcsname{#7}\addcontentsline
    {toc}{#1}{\ifnum #2>\c@secnumdepth \else
     \protect\numberline{\csname the#1\endcsname}\fi
           #7}\else
    \def\@svsechd{#6\hskip #3\@svsec #8\csname #1mark\endcsname
          {#7}\addcontentsline
          {toc}{#1}{\ifnum #2>\c@secnumdepth \else
     \protect\numberline{\csname the#1\endcsname}\fi
           #7}}\fi
     \@xsect{#5}}
\def\label#1{\@bsphack\if@filesw {\let\thepage\relax
   \xdef\@gtempa{\write\@auxout{\string
   \newlabel{#1}{{\thesection.\@currentlabel}{\thepage}}}}}\@gtempa
   \if@nobreak \ifvmode\nobreak\fi\fi\fi\@esphack}
\def\@eqnnum{(\thesection.\theequation)}
\def\section{\setcounter{equation}{0} \@startsection {section}{1}{\z@}{-3.5ex
   plus -1ex minus -.2ex}{2.3ex plus .2ex}{\Large\bf}}
\newcount\@minsofar
\newcount\@min
\newcount\@cite@temp
\def\@citex[#1]#2{%
\if@filesw \immediate \write \@auxout {\string \citation {#2}}\fi
\@tempcntb\m@ne \let\@h@ld\relax \def\@citea{}%
\@min\m@ne%
\@cite{%
  \@for \@citeb:=#2\do {\@ifundefined {b@\@citeb}%
    {\@h@ld\@citea\@tempcntb\m@ne{\bf ?}%
    \@warning {Citation `\@citeb ' on page \thepage \space undefined}}%
{\@minsofar\z@ \@for \@scan@cites:=#2\do {%
  \@ifundefined{b@\@scan@cites}%
    {\@cite@temp\m@ne}
    {\@cite@temp\number\csname b@\@scan@cites \endcsname \relax}%
\ifnum\@cite@temp > \@min% select the next one to list
    \ifnum\@minsofar = \z@
      \@minsofar\number\@cite@temp
      \edef\@scan@copy{\@scan@cites}\else
    \ifnum\@cite@temp < \@minsofar
      \@minsofar\number\@cite@temp
      \edef\@scan@copy{\@scan@cites}\fi\fi\fi}\@tempcnta\@min
  \ifnum\@minsofar > \z@ % some more
    \advance\@tempcnta\@ne
    \@min\@minsofar
    \ifnum\@tempcnta=\@minsofar %   Number follows previous--hold on to it
      \ifx\@h@ld\relax
        \edef \@h@ld{\@citea\csname b@\@scan@copy\endcsname}%
    \else \edef\@h@ld{\ifmmode{-}\else--\fi\csname b@\@scan@copy\endcsname}%
      \fi
    \else \@h@ld\@citea\csname b@\@scan@copy\endcsname
          \let\@h@ld\relax
  \fi % no more
\fi}%
\def\@citea{,\penalty\@highpenalty\,}}\@h@ld}{#1}}
\def\appendixname{Appendix}
\def\appendix{\par
  \def\pre@section{\appendixname{}}
  \setcounter{section}{1}
  \@addtoreset{equation}{section}
  \def\thesection{\Alph{section}}
  \def\theequation{\arabic{equation}}}

\makeatother

\begin{document}
\addtolength{\unitlength}{-0.5\unitlength}
\newsavebox{\near}\savebox{\near}(30,30){\begin{picture}(30,30)
\thicklines\put(10,20){\line(1,-1){10}}\put(10,20){\line(2,1){20}}
\put(20,10){\line(1,2){10}}\end{picture}}

\newsavebox{\nwar}\savebox{\nwar}(30,30){\begin{picture}(30,30)
\thicklines\put(10,10){\line(1,1){10}}\put(10,10){\line(-1,2){10}}
\put(20,20){\line(-2,1){20}}\end{picture}}

\newsavebox{\swar}\savebox{\swar}(30,30){\begin{picture}(30,30)
\thicklines\put(10,20){\line(1,-1){10}}\put(0,0){\line(2,1){20}}
\put(0,0){\line(1,2){10}}\end{picture}}

\newsavebox{\sear}\savebox{\sear}(30,30){\begin{picture}(30,30)
\thicklines\put(10,10){\line(1,1){10}}\put(10,10){\line(2,-1){20}}
\put(20,20){\line(1,-2){10}}\end{picture}}

\newsavebox{\nar}\savebox{\nar}(30,30){\begin{picture}(30,30)
\thicklines\put(5,10){\line(1,0){20}}\put(5,10){\line(1,2){10}}
\put(15,30){\line(1,-2){10}}\end{picture}}

\newsavebox{\war}\savebox{\war}(30,30){\begin{picture}(30,30)
\thicklines\put(0,15){\line(2,1){20}}\put(0,15){\line(2,-1){20}}
\put(20,5){\line(0,1){20}}\end{picture}}

\newsavebox{\sar}\savebox{\sar}(30,30){\begin{picture}(30,30)
\thicklines\put(15,0){\line(1,2){10}}\put(15,0){\line(-1,2){10}}
\put(5,20){\line(1,0){20}}\end{picture}}

\newsavebox{\ear}\savebox{\ear}(30,30){\begin{picture}(30,30)
\thicklines\put(10,5){\line(0,1){20}}\put(10,5){\line(2,1){20}}
\put(10,25){\line(2,-1){20}}\end{picture}}

\newsavebox{\mybox}\savebox{\mybox}(350,350){
\begin{picture}(350,350)
\thicklines
\multiput(30,112.5)(15,0){16}{\line(1,0){10}}
\multiput(30,187.5)(15,0){16}{\line(1,0){10}}
\multiput(112.5,30)(0,15){16}{\line(0,1){10}}
\multiput(187.5,30)(0,15){16}{\line(0,1){10}}
\multiput(75,150)(75,75){2}{\circle*{15}}
\multiput(150,75)(75,75){2}{\circle*{15}}
\multiput(74,150)(1,0){3}{\line(1,1){75}}
\multiput(74,150)(1,0){3}{\line(1,-1){75}}
\multiput(149,75)(1,0){3}{\line(1,1){75}}
\multiput(149,225)(1,0){3}{\line(1,-1){75}}
\multiput(134,210)(1,0){3}{\line(-2,-1){20}}
\multiput(136,210)(0,1){3}{\line(-1,-2){10}}
\multiput(209,165)(1,0){3}{\line(-2,1){20}}
\multiput(211,165)(0,-1){3}{\line(-1,2){10}}
\multiput(134,90)(1,0){3}{\line(-2,1){20}}
\multiput(136,90)(0,-1){3}{\line(-1,2){10}}
\multiput(166,90)(-1,0){3}{\line(2,1){20}}
\multiput(164,90)(0,-1){3}{\line(1,2){10}}
\end{picture}}

\newsavebox{\stars}\savebox{\stars}(350,350){
\begin{picture}(350,350)
\thicklines
\multiput(40,112.5)(15,0){16}{\line(1,0){10}}
\multiput(40,187.5)(15,0){16}{\line(1,0){10}}
\multiput(187.5,30)(0,15){16}{\line(0,1){10}}
\multiput(112.5,30)(0,15){16}{\line(0,1){10}}
\multiput(75,225)(150,0){2}{\circle*{15}}
\multiput(75,75)(150,0){2}{\circle*{15}}
\put(150,150){\circle*{15}}
\multiput(149,150)(1,0){3}{\line(1,1){75}}
\multiput(149,150)(1,0){3}{\line(1,-1){75}}
\multiput(74,75)(1,0){3}{\line(1,1){75}}
\multiput(74,225)(1,0){3}{\line(1,-1){75}}
\multiput(110,160)(75,-75){2}{\usebox{\sear}}
\put(110,110){\usebox{\near}}\put(160,160){\usebox{\swar}}
\end{picture}}

\newsavebox{\starss}\savebox{\starss}(350,350){
\begin{picture}(350,350)
\thicklines
\multiput(40,112.5)(15,0){16}{\line(1,0){10}}
\multiput(40,187.5)(15,0){16}{\line(1,0){10}}
\multiput(187.5,30)(0,15){16}{\line(0,1){10}}
\multiput(112.5,30)(0,15){16}{\line(0,1){10}}
\multiput(75,225)(150,0){2}{\circle*{15}}
\multiput(75,75)(150,0){2}{\circle*{15}}
\put(150,150){\circle*{15}}
\multiput(149,150)(1,0){3}{\line(1,1){75}}
\multiput(149,150)(1,0){3}{\line(1,-1){75}}
\multiput(74,75)(1,0){3}{\line(1,1){75}}
\multiput(74,225)(1,0){3}{\line(1,-1){75}}
\multiput(110,160)(75,-75){2}{\usebox{\sear}}
\put(185,185){\usebox{\near}}\put(85,85){\usebox{\swar}}
\end{picture}}

\def\b{\beta}
\def\d{\delta}
\def\g{\gamma}
\def\a{\alpha}
\def\s{\sigma}
\def\l{\lambda}
\def\e{\epsilon}
\def\r{\rho}
\def\d{\delta}
\def\wid{\widehat}
\def\ds{\displaystyle}
\def\be{\begin{equation}}
\def\ee{\end{equation}}
\def\beq{\begin{eqnarray}}
\def\eeq{\end{eqnarray}}
\def\ov{\overline}
\def\om{\omega}
\rightline{\large September, 1998}
\vspace{2cm}
\centerline{\LARGE Functional relations and nested Bethe ansatz}
\centerline{\LARGE for $sl(3)$ chiral Potts model at $q^2=-1$}.
\vspace{1cm}

\centerline{\Large H.E. Boos\footnote{
On leave of absence 
from Institute for High Energy Physics, Protvino, 142284, Russia,
E-mail: boos@avzw02.physik.uni-bonn.de}}
\centerline{\large
Physikalisches Institut der Universit{\"a}t Bonn}
\centerline{\large 53115, Bonn,
Germany}

\vspace{1cm}
\centerline{\Large V.V. Mangazeev\footnote{E-mail:
vladimir@maths.anu.edu.au}}
\centerline{\large Centre for Mathematics and its Applications,}
\centerline{\large  School of Mathematical Sciences,}
\centerline{\large The Australian National University,}
\centerline{\large Canberra, ACT 0200, Australia}

\vspace{1cm}

\begin{abstract}
We obtain the functional relations for the eigenvalues of
the transfer matrix of the $sl(3)$ chiral Potts model for
$q^2=-1$. For the homogeneous model in both directions a solution
of these functional relations can be written in terms of roots
of Bethe ansatz-like equations. In addition, a direct nested
Bethe ansatz has also been  developed for this case. 
\end{abstract}

\section{Introduction}

The discovery of the  chiral Potts model (CPM)  and new $N$-state 
solutions of the Yang-Baxter relation \cite{YMP,MPTS,BPY,YP}
is one of the most impressive results in the theory of two-dimensional
lattice integrable systems \cite{BaxB}. A remarkable feature of this
model is that the spectral parameters belong to a high genus
algebraic curve. The reason of this fact remained unclear until
it was shown that the weights of the 
chiral Potts model naturally appear as intertwiners of the
cyclic representations for $sl(2)$ $L$-operators \cite{BS} related to
the six-vertex model \cite{L}. 

The next step was to generalise the chiral Potts model for the $sl(n)$ case
for $q^N=-1$ \cite{BKMS}. As a result, a new $N^{n-1}$-state family
of generalised chiral Potts models has been obtained with Boltzmann 
weights satisfying the Yang-Baxter relation \cite{DJMM,KMN}. 

Then Bazhanov and Baxter \cite{BB} made a remarkable
observation that the $sl(n)$ chiral Potts model with $N=2$ is related
to the integrable three-dimensional Zamolodchikov model \cite{Z1,Z2}.
Therefore, a new link between two quite different 
regions of the theory of integrable models has been revealed.  
It is well known that the tetrahedron equations provide 
sufficient conditions for a commutativity of the transfer matrices
in three dimensions \cite{BS1,JM}. 
As a result, a new $N$-state three-dimensional family 
of integrable models has been discovered with Boltzmann weights
satisfying the tetrahedron equation \cite{BT,KMS}. 

The Zamolodchikov model attracted a lot of attention 
even before the appearance of the chiral Potts model. Its partition
function in the thermodynamic limit has been calculated in \cite{Bax} 
and the related hamiltonian has been  studied in \cite{BQ}. 
It was shown in \cite{BQ} that the Zamolodchikov model with two layers is 
equivalent to the critical two-dimensional free-fermion model.
However, the structure of the spectrum of Zamolodchikov model
with three and more layers remains unclear. 

In this paper we try to take several steps towards the understanding of 
the structure of the spectrum of $sl(3)$ chiral Potts model which
is equivalent to the ``modified'' (in the terminology of \cite{Bax,BB})
three-layer Zamolodchikov model. Starting from the general case
we carefully analyse the homogeneous case of the $sl(3)$
chiral Potts model where a usual nested Bethe ansatz technique
\cite{V} can be applied.

The paper is organised as follows. In Section 2 we give a 
full description of the $sl(3)$ chiral Potts model and introduce
all necessary notations. In Section 3 we describe a ``fusion''
procedure \cite{KRS,KR,BR} for the $sl(3)$ chiral Potts model
for $q^2=-1$ ($N=2$) and obtain the closed functional equations
for eigenvalues of the transfer matrix. Section 4 is devoted
to the spectrum of the model in the completely homogeneous case. 
In Section 5 we apply the nested Bethe ansatz technique 
to obtain eigenvalues of the
transfer matrix for the homogeneous case.
In the last section we discuss our
results and further directions for investigation. 

\section{The $sl(3)$ chiral Potts model}

In this section we give all necessary definitions for the
$sl(3)$ chiral Potts model with fixed $N$. 
All definitions and
formulas of this section are just a specification of the $sl(n)$ case
for $n=3$ \cite{KMN}. 

Consider an oriented square lattice ${\cal L}$ and
its medial lattice ${\cal L}'$ (shown in Figure 1
by solid and dashed lines, respectively).
The oriented vertical (horizontal) lines of  ${\cal L}'$ carry 
rapidity variables $p_1,p_2$ ($r_1,r_2$) in an alternating order
(note that the orientations of rapidity lines shown by open
arrows alternate too). The edges of the lattice 
${\cal L}$ are oriented in such a way that all the NW-SE edges have the
same (NW-SE) direction while the NE-SW edges are oriented
in a checkerboard order.

Each rapidity variable is represented by three $2$-vectors 
($h_i^+(p)$, $h_i^-(p)$), $i=1,2,3$
which specify the point $p$ of the algebraic curve $\Gamma$ 
defined by relations
\be
\pmatrix{h_i^+(p)^N\cr h_i^-(p)^N\cr}=K_{ij}
\pmatrix{h_j^+(p)^N\cr h_j^-(p)^N\cr},\quad \forall i,j=1,2,3.  \label{l1}
\ee
where $K_{ij}$ are $2\times2$ complex matrices of moduli satisfying
relations
\be
detK_{ij}=1,\quad K_{ii}=K_{ij}K_{jk}K_{ki}=1.    \label{l2}
\ee
Hereafter we imply that indices $i,j,k,\ldots$ run the
values $1,2,3$ {\it modulo} $3$.

Below we need the automorphism $\tau$ on the 
curve $\Gamma$ defined as follows
\be
h_j^+(\tau(p))=h_j^+(p),\quad
h_j^-(\tau(p))=\om h_j^-(p), \quad j=1,2,3. \label{l2b}
\ee

On each site of the lattice ${\cal L}$ place $Z_N\times Z_N$ spins,
which are described by a local variable
\be
\a=(\a_1,\a_2),\quad \a_1,\a_2=0,1,\ldots,N-1.   \label{l3}
\ee

There are two kinds of neighbouring local state pairs
depending on the relative orientation of the dashed and solid lines
as indicated in Figure~2, with states $\a$ and $\b$, and Boltzmann weights
$\overline W_{pq}(\a,\b)$ and  $(\overline W_{qp}(\a,\b))^{-1}$ on the
edges of  ${\cal L}$.

The function $\ov W_{pq}(\a,\b)$, $\a,\b\in Z_N\times Z_N$ is defined 
by the following relations
\be
\ov W_{p,q}(\a,\b)=\om^{Q(\a,\b)}g_{pq}(0,\a-\b), \quad
\om=\exp({2\pi i\over N}), \label{l5}
\ee
where
\be
Q(\a,\b)=\b_1(\b_1-\a_1)+\b_2(\b_1-\a_1+\b_2-\a_2),
\>\a,\b\in Z_N\times Z_N\label{l6}
\ee
and the function $g_{pq}(0,\a)$ has the following form
\be
g_{pq}(0,\a)={\ds\prod_{\b=0}^{\a_1+\a_2-1}(h_0^+(p)h_0^-(q)-h_0^+(q)h_0^-(p)
\om^{-\b})\over  \ds
\prod_{i=1}^2
\prod_{\b_i=0}^{\a_i-1}(h_i^+(p)h_i^-(q)-h_i^+(q)h_i^-(p)\om^{1+\b_i})}
\label{l7}
\ee

We choose a normalisation of $\ov W_{pq}(\a,\b)$ as 
\be
\ov W_{pq}(0,0)=1. \label{l8}
\ee
Then it is easy to see that
\be
\ov W_{pp}(\a,\b)=\ov\d_{\a,\b},\quad
    \overline{\delta}_{\a,\b}\equiv\cases{1,&$\a=\b\pmod{N};$\cr
                                           0,& otherwise.\cr}  \label{l10}
\ee
The function $\ov W_{pq}(\a,\b)$ satisfies the inversion relation
\be
\sum_{\b\in Z_2\times Z_2}\overline W_{pq}(\a,\b)\overline W_{qp}(\b,\g)=
                \overline\delta_{\a,\g}\Phi_{pq},                 \label{l11}
\ee
where
\be
\Phi_{pq}=N^2{x_p^N-x_q^N\over x_p-x_q}\prod_{i=1}^3
{x_i(p)-x_i(q)\over x_i(p)^N-x_i(q)^N},\quad 
x_i(p)\equiv {h_i^-(p)\over h_i^+(p)},\quad 
x_p\equiv\prod_{i=1}^3x_i(p).                             \label{l12}
\ee

Now let us suppose that our lattice ${\cal L}$ has $M$ sites in 
a horizontal direction and $L$ sites in a vertical one ($L$ should be even).
As usual we imply cyclic boundary conditions in both
directions.

Let us denote the spin variables of three consecutive   rows
as $\s_1,\ldots,\s_M$, $\s'_1,\ldots,\s'_M$ and $\s''_1,\ldots,s''_M$
(see Figure 1). Then we can define two $N^{2M}\times N^{2M}$ 
transfer matrices $T_{p_1}$ and $\ov T_{p_2}$ of the length $M$
\be
[T_{p_1}]_{\s_1\ldots\s_M}^{\s'_1\ldots\s'_M}=
\prod_{\i=1}^M\ov W_{p_1r_1}(\s_i,\s'_i)
\ov W_{r_2p_1}(\s'_i,\s_{i+1}),\label{l13}
\ee
\be
[\ov T_{p_2}]_{\s'_1\ldots\s'_M}^{\s''_1\ldots\s''_M}=
\prod_{\i=1}^M{\ov W_{p_2r_2}(\s''_{i+1},\s'_i)\over
\ov W_{p_2r_1}(\s''_i,\s'_i)} \label{l15}
\ee
Let us fix the rapidity variables $r_1$, $r_2$ and use more simple
notations $T_{p_1}$ and $\ov T_{p_2}$.

%Hereafter we will show by subscripts a dependence of 
%the transfer-matrices $T_{p_1}$ and $\ov T_{p_2}$ on the spectral parameters 
%$p_1$, $p_2$ and regard that the rapidity variables $r_1$, $r_2$
%are fixed.

The partition function is
\be
Z= Tr(T_{p_1}\ov T_{p_2})^{L/2}. \label{l16}
\ee

In the next sections we derive several functional relations
between $T_{p_1}$ and $\ov T_{p_2}$.

%In the next sections we will exhibit several functional relations
%between $T_{p_1}$ and $\ov T_{p_2}$ 
%at $N=2$ which should enable
%their eigenvalues to be calculated.

We can construct two $R$-matrices from the weights $\ov W_{pq}(\a,\b)$.
 Define (see Figure~3)
\be
{\ds\ov S_{\g,\a}^{\d,\b}(p_1,p_2;r_2,r_1)}
{=\ds{\ov W_{r_2p_1}(\g,\a)\ov W_{p_1r_1}(\a,\d)\ov W_{p_2r_2}(\b,\g)\over
\ov W_{p_2r_1}(\b,\d)}}        \label{l17}
\ee
This $R$-matrix satisfies the Yang-Baxter equation
\beq
&{\ds \ov S_{12}(p_1,p_2;q_1,q_2)\ov S_{13}(p_1,p_2;r_1,r_2)
\ov S_{23}(q_1,q_2;r_1,r_2)=}&\nonumber\\
&{\ds=\ov S_{23}(q_1,q_2;r_1,r_2)
\ov S_{13}(p_1,p_2;r_1,r_2)\ov S_{12}(p_1,p_2;q_1,q_2).}& \label{l18}
\eeq

It leads us to the following commutation relation for the product
of the transfer matrices $\ov T_{p_2}T_{p_1}$
\be
\ov T_{p_2}T_{p_1}\ov T_{q_2}T_{q_1}=\ov T_{q_2}T_{q_1}
\ov T_{p_2}T_{p_1},  \label{l19}
\ee
where $p_1$, $p_2$, $q_1$, $q_2$ are four points on the curve $\Gamma$.

Similarly one can introduce the second $R$-matrix $S$ (see \cite{KMN}, for
example)
and prove another commutation relation
\be
T_{p_1}\ov T_{p_2}T_{q_1}\ov T_{q_2}=T_{q_1}\ov T_{q_2}
T_{p_1}\ov T_{p_2}.  \label{l20}
\ee

However, all commutation relations between the transfer matrices are the 
consequence of the only equation which is called "star-star" 
relation \cite{K},\cite{BB}. 
To write it down define two "stars" (see Figures 4,5)
\be
W_{r_2r_1}^{p_1p_2}(\a,\b,\g,\d)=\sum_\s
{\ov W_{p_1r_1}(\a,\s)\ov W_{p_2r_2}(\g,\s)\ov W_{r_2p_1}(\s,\b)\over
\ov W_{p_2r_1}(\d,\s)}                        \label{l21}
\ee
and
\be
\widetilde W_{r_2r_1}^{p_2p_1}(\a,\b,\g,\d)=\sum_\s
{\ov W_{p_1r_1}(\s,\g)\ov W_{p_2r_2}(\s,\a)\ov W_{r_2p_1}(\d,\s)\over
\ov W_{p_2r_1}(\s,\b)}.                        \label{l22}
\ee

In these notations the star-star relation can be written as 
(see Figure 6)
\beq
&\ov W_{p_2p_1}(\d,\a)\ov W_{r_2r_1}(\d,\g)W_{r_1r_2}^{p_1p_2}
(\a,\b,\g,\d)=&\nonumber\\
&\widetilde W_{r_2r_1}^{p_2p_1}(\a,\b,\g,\d)\ov W_{r_2r_1}(\a,\b)
\ov W_{p_2p_1}(\g,\b).&
\label{l23}
\eeq
The Yang-Baxter equation (\ref{l18}) can be easily obtained
by repeated applications of (\ref{l23}).

Now using (\ref{l23}) we obtain one more "commutation"
relation between the transfer matrices $T_{p_1}$, $\ov T_{p_2}$.
To do that define the shift operator 
\be
X_{\s_1\ldots\s_M}^{\s'_1\ldots\s'_M}=\prod_{i=1}^M\ov\d_{\s_i,\s'_{i+1}}
\label{l24}
\ee
Then using cyclic boundary conditions in a horizontal direction and 
relation (\ref{l23}) we obtain 
\be
T_{p_1}\ov T_{p_2} U^{(d)} X^{-1}=
U^{(d)} X^{-1} \ov T_{p_2} T_{p_1}, \label{l25}
\ee
where $U^{(d)}$ is the diagonal matrix independent of $p_1$, $p_2$
\be
[U^{(d)}]^{\s'_1,\ldots,\s'_M}_{\s_1,\ldots,\s_M}=
\prod_{i=1}^M\ov\d_{\s_i,\s'_i} \ov W_{r_2,r_1}(\s_i,\s_{i+1}). \label{l26}
\ee

\section{Functional relations at $q^2=-1$}.

Here we consider only the case $q^2\equiv\om=-1$.
The product of the transfer matrices $T_{p_1}\ov T_{p_2}$ can
be easily rewritten in terms of ``star'' weights (\ref{l21})
\be
T_{p_1}\ov T_{p_2}=\prod_{i=1}^M W_{r_1r_2}^{p_1p_2}(\s_i,\s_{i+1},
\s'_{i+1},\s'_i).          \label{f1}
\ee
We can specify the horizontal rapidities in such a way that
the left and right ``gauge'' factors $\ov W_{p_1,p_2}(\a,\b)$ in (\ref{l23})
become degenerate.
Namely, there are two simplest choices to do that
\be
p_2=\tau^\l(p),\quad p_1=p, \quad \l=0,1, \label{f2}
\ee
where  $\tau$ is defined in (\ref{l2b}) and
the case $\l=0$ corresponds just to $p_2=p_1=p$.

First let us set $p_2=p_1=p$. 
Then using (\ref{l23}) and the explicit form of $\ov W_{p_1p_2}(\a,\b)$
it is easy to see that we have
the only non-zero matrix elements for $W^{pp}_{r_1r_2}(\a,\b,\g,\a)$ 
if $\b=\g$. 
Similarly setting $p_2=\tau(p_1)$ we obtain that the only
non-zero matrix elements for $\ov W^{p\tau(p)}_{r_1r_2}(\a,\b,\b,\d)$
are when $\a=\d$.

Therefore, we obtain  for
$T_{p}\ov T_{\tau^\l(p)}$ that 
if $\s_I=\s''_{I}$ for some $I=1,\ldots,M$ then
\be
[T_p\ov T_{\tau^\l(p)}]_{\s_1,\ldots,\s_M}^{\s''_1,\ldots,\s''_M}=0, 
\quad\mbox{if}
\quad \s_J\neq\s''_J \quad\mbox{for some}\quad J\neq I,\quad J=1,\ldots,M
\label{f3}
\ee
As a result we can split $T_{p}\ov T_{\tau^\l(p)}$ into 
``diagonal'' and ``non-diagonal'' parts. To do this explicitly
consider two $4\times4$ projectors
\be
P^+_{\g,\d}={1\over4},\quad P^-_{\g,\d}=\d_{\g,\d}-{1\over4},
\quad \g,\d=1,\ldots,4.
\label{f4}
\ee
Now let us consider $R$-matrix (\ref{l17}) as $4\times4$ matrix with
respect to horizontal indices with fixed vertical ones
and denote it as $\ov S_{p_1p_2}(\a,\b)$. Then we have
\be  
P^+\ov S_{pp}(\a,\b)P^+=P^+\ov S_{pp}(\a,\b),\quad 
P^-\ov S_{pp}(\a,\b)P^-=\ov S_{pp}(\a,\b)P^-, \label{f5}
\ee
\be 
P^+\ov S_{p\tau(p)}(\a,\b)P^+=
\ov S_{p\tau(p)}(\a,\b)P^+,\quad
P^-\ov S_{p\tau(p)}(\a,\b)P^-=P^-\ov S_{p\tau(p)}
(\a,\b). \label{f7}
\ee
As a consequence we have the following decomposition of the product
$T_p\ov T_{\tau^\l(p)}$
\be
T_p\ov T_{\tau^\l(p)} = 
Tr(P^+\ov S_{p\tau^\l(p)}(\a,\b)P^+)^M+
Tr(P^-\ov S_{p\tau^\l(p)}(\a,\b)P^-)^M.     \label{f8}
\ee
It is not difficult to check that
\be
Tr(P^+\ov S_{p\tau^\l(p)}(\a,\b)P^+)^M=\Phi_{p,r_{\l+1}}^MI,\label{f8a}
\ee
where $I$ being the identity matrix $4^M\times4^M$,
$\Phi_{pq}$ is defined in (\ref{l12}) and 
all subscripts should be considered ${\it modulo}$ $2$.
Denote the second term in (\ref{f8}) as
\be
T^{(\l)}(p;r_{1+\l},r_{2+\l})=Tr(P^-\ov S_{p\tau^\l(p)}(\a,\b)P^-)^M. \label{f9}
\ee
The transfer matrix (\ref{f9}) has nonzero matrix elements only
if $\s_I\neq\s''_I$, $\forall I=1,\ldots,M$.

It is not difficult to construct local "fused" weights for the
transfer matrix $T^{(\l)}(p;r_{1+\l},r_{2+\l})$. 
First let us write the following decomposition for 
$P^-$
\be
P_{\g,\d}^-=\sum_{i=1}^3c(i,\g)c(i,\d),\quad
c(i,\a)={1\over2}\om^{i\a_1+{1\over2}i(i-1)\a_2},\quad i=1,2,3,\>
\a\in Z_2\times Z_2.\label{f10a}
\ee
Now consider the family
of "fused" $L$-operators which act in the tensor product
of the auxiliary space $C^3$ and the quantum space $C^4$
\be
L^{(\l)}_{ij}(p;r_1,r_2)=\Lambda_{p;r_1,r_2}\sum_{\g,\d\in Z_2\times Z_2}
c(i,\g)c(j,\d)\ov S_{\g,\a}^{\d,\b}(p,\tau^\l(p);r_2,r_1),
 \label{f11}
\ee
\be
\Lambda_{p;r_1,r_2}={1\over4}
{h_3^-(r_1)^2h_3^+(p)^2-h_3^-(p)^2h_3^+(r_1)^2\over
 h_3^-(r_2)^2h_3^+(p)^2-h_3^-(p)^2h_3^+(r_2)^2}
 \prod_{i=1}^3(h_i^-(r_2)h_i^+(p)+h_i^-(p)h_i^+(r_2))  \label{f12}
\ee
and $\ov S_{\g,\a}^{\d,\b}(p_1,p_2;r_2,r_1)$ is defined in (\ref{l17}).

Introduce the transfer matrix $t^{(\l)}(p;r_1,r_2)$ constructed
from the ``fused'' $L$-operators $L^{(\l)}_{ij}(p;r_1,r_2)$ over the auxiliary
space $C^3$
\be
t^{(\l)}(p;r_1,r_2)_{\s_1,\ldots,\s_M}^{\s''_1,\ldots,\s''_M}=
\sum_{i_1}\ldots\sum_{i_M}
\prod_{\a=1}^M [L^{(\l)}_{i_\a,i_{\a+1}}
(p;r_1,r_2)]_{\s_\a}^{\s''_\a}. \label{f13}
\ee
It is easy to check that two transfer matrices $T^{(\l)}(p;r_1,r_2)$ in 
(\ref{f9}) and $t^{(\l)}(p;r_1,r_2)$ are related as follows
\be
T^{(\l)}(p;r_1,r_2)={1\over \Lambda_{p;r_1,r_2}^M}t^{(\l)}(p;r_1,r_2).  
\label{f14}
\ee

Therefore, the problem of calculating eigenvalues for the
product of two transfer matrices $T_p\ov T_{\tau^\l(p)}$ in (\ref{f8}) 
is reduced to a calculation of the eigenvalues for
the transfer matrix $t^{(\l)}(p;r_1,r_2)$ in (\ref{f13}).

Let us give explicit formulas for the matrix elements
of $L^{(\l)}_{ij}(p;r_1,r_2)$. We have
\be
L^{(\l)}_{ii}(p;r_1,r_2)=u_{i+1}^+(p;r_1,r_2)X_{i+1}+
u_{i-1}^-(p;r_1,r_2)X_{i-1}, 
\label{f15}
\ee
\be
L^{(\l)}_{i,i+1}(p;r_1,r_2)=Z_{i,i+1}(v^{(\l)}_{i+1-\l}(p;
r_1,r_2)X_{i+1-\l}+w_{i-1}^{(1-\l)}(p;r_1,r_2)
X_{i-1}),\quad \label{f16}                  
\ee
\be
L^{(\l)}_{i,i-1}(p;r_1,r_2)=Z_{i,i-1}(
v^{(1-\l)}_{i-1+\l}(p;r_1,r_2)X_{i-1+\l}+
w^{(\l)}_{i+1}(p;r_1,r_2)X_{i+1}), \label{f17}               
\ee
where
\be
u^\pm_{i}(p;r_1,r_2)=
{h_i^\pm(r_2)\over h_i^\pm(r_1)}
\prod_{\a=1}^3h_\a^\mp(p)h_\a^\pm(r_1),\label{f18}
\ee
\be
v_i^{({0\atop1})}(p;r_1,r_2)=
h_i^\mp(p)h_i^\pm(r_2)h_{i+1}^-(r_1)h_{i+1}^+(p)h_{i-1}^-(p)h_{i-1}^+(r_1)
\label{f19}
\ee
\be
w_i^{({0\atop1})}(p;r_1,r_2)=
h_i^\mp(p)h_i^\pm(r_2)h_{i+1}^\mp(r_1)h_{i+1}^\pm(p)
h_{i-1}^\mp(r_1)h_{i-1}^\pm(p),
\label{f20}
\ee
the $4\times4$ matrices $X_i$ and $Z_{i,j}$ act in $Z_2\times Z_2$
and have the following matrix elements
\be
<\a|X_1|\b>=\d_{\a_1,\b_1}\d_{\a_2,\b_2-1}, \quad
<\a|X_2|\b>=\d_{\a_1,\b_1-1}\d_{\a_2,\b_2},    \label{f21}
\ee
\be
<\a|Z_1|\b>=\om^{\a_2}\d_{\a_1,\b_1}\d_{\a_2,\b_2},\quad 
<\a|Z_2|\b>=\om^{\a_1}\d_{\a_1,\b_1}\d_{\a_2,\b_2}, \label{f22}
\ee
\be
X_1X_2X_3=1,\quad Z_3=1, \quad Z_{ij}=Z_iZ_j^{-1}, \quad
i\neq j=1,2,3.\label{f23}
\ee
They satisfy usual relations
\be
X_iZ_{jk}=\om^{\d_{ij}-\d_{ik}}Z_{jk}X_i,\quad Z_{ij}Z_{jk}Z_{ki}=1.
\label{f24}
\ee

In fact, it is convenient to introduce the ``gauge-transformed'' 
$L$-operators 
\be
{\ds\widehat L^{(\l)}_{ij}(x_p;r_1,r_2)=
{\Bigl[\ds\prod_{i=1}^3h_i^+(p)\Bigr]^{-1}}[\xi_i(p)]^{(-1)^\l}
L^{(\l)}_{ij}(p;r_1,r_2)[\xi_j(p)]^{(-1)^{1+\l}}},
\label{f25}
\ee
where
\be
\xi(p)\equiv(\xi_1(p),\xi_2(p),\xi_3(p))=(x_3(p),1,x_1^{-1}(p)). \label{f26}
\ee
and $x_p$, $x_i(p)$ are defined in (\ref{l13}).

It is easy to see that $L$-operators (\ref{f25}) depend on the 
rapidity variable $p$ only via the combination $x_p$. Therefore,
we have
\be
t^{(\l)}(p;r_1,r_2)=(\prod_{i=1}^3 h_i^+(p))^M t^{(\l)}(x_p;r_1,r_2), 
\label{f27}
\ee
where $t^{(\l)}(x_p;r_1,r_2)$ is the
transfer matrix constructed from the $L$-operators (\ref{f25}) 
and  obviously just a polynomial in the variable $x_p$ of the
degree $M$. 

It follows that we can consider the parameters $x_p$, $h_i^\pm(r_1)$,
$h_i^\pm(r_2)$ as independent ones and do not take into account
the surface equations (\ref{l1}).  

$L$-operators (\ref{f25}) satisfy the following Yang-Baxter equation
\be
{\ds\sum_{i_2,j_2}
R_{i_1,i_2;j_1,j_2}^{(\l)}(x/y)\wid L^{(\l)}_{i_2i_3}(x^2)
\wid L^{(\l)}_{j_2j_3}(y^2)}{\ds=\sum_{i_2,j_2}\wid L^{(\l)}_{j_1j_2}(y^2)
\wid L^{(\l)}_{i_1i_2}(x^2)
R_{i_2,i_3;j_2,j_3}^{(\l)}(x/y)}, \label{f28} 
\ee
where we omit a dependence on $r_1,r_2$ in $\wid L^{(\l)}(x;r_1,r_2)$ and
$R_{i_1,i_2;j_1,j_2}^{(\l)}(x)$ coincides with the deformed
trigonometric $sl(3)$ $R$-matrix. 
Namely, let us assume for a moment that $q$ is arbitrary
and introduce $R$-matrix \cite{CKP}
\beq
&R_{i_1,i_2;j_1,j_2}(x,q,\rho)=
\d_{i_1i_2}\d_{j_1j_2}\d_{i_1j_1}(q-1)(x+x^{-1}q^{-1})+&\nonumber\\
&+\d_{i_1i_2}\d_{j_1j_2}\rho_{i_1j_1}(x-x^{-1})+
\d_{i_1j_2}\d_{i_2j_1}(1-\d_{i_1j_1})\s_{i_1i_2}(x),&  \label{f29}
\eeq
where
\be
\rho_{ii}=\rho_{ij}\rho_{ji}=1,\quad
\sigma_{ij}\equiv\cases{0,               &$i=j;$\cr
                         (q-q^{-1})x,     &$i<j;$\cr
                         (q-q^{-1})x^{-1},&$i>j.$\cr}      \label{f31}
\ee
Then
\be
R_{i_1,i_2;j_1,j_2}^{(\l)}(x)=R_{i_1,i_2;j_1,j_2}(x^{(-1)^\l},q,\rho)
\label{f32}
\ee
with
\be
q=i,\quad
\rho_{j,j+1}=\rho_{j+1,j}^{-1}=i^{(-1)^{\l+1}},\quad j=1,2,3. \label{f33}
\ee
Now using a fusion technique for the $R$-matrix (\ref{f29}) we can 
construct functional equations for the transfer matrix
$t^{(\l)}(x_p;r_1,r_2)$ .
Actually this procedure is quite similar to the fusion for the $sl(3)$
trigonometric $R$-matrix, so we do not give a detailed derivation
of the functional equations. 

We can define $L$-operators related to the "antisymmetric" 
representation
\beq
&\ov L^{(\l)}_{ij}(x;r_1,r_2)=
{\ds{\ds1\over\phi_1^-(x;r_{1+\l})}}
\Bigl[
\wid L_{km}^{(\l)}((-1)^\l x;r_1,r_2)
\wid L_{ln}^{(\l)}((-1)^{1+\l}x;r_1,r_2)-&\nonumber\\
&-(-1)^{\l+\d_{i,2}}
\wid L_{lm}^{(\l)}((-1)^\l x;r_1,r_2)
\wid L_{kn}^{(\l)}((-1)^{1+\l}x;r_1,r_2)\Bigr],&\label{f34}
\eeq
where indices $\{i,k,l\}$ and $\{j,m,n\}$ are {\bf even} permutations
of $\{1,2,3\}$ and 
we define
\be
\phi_1^\pm(x;r)=
(\prod_{i=1}^3h_i^-(r)\pm x\prod_{i=1}^3h_i^+(r)). \label{f35}
\ee
Let $\ov t^{(\l)}(x;r_1,r_2)$ be the transfer matrix constructed from
the $L$-operators (\ref{f34}).

Omitting a dependence of $t^{(\l)}(x_p;r_1,r_2)$, 
$\ov t^{(\l)}(x_p;r_1,r_2)$ on 
 $r_1$, $r_2$ one can  show that the following system of functional
 equations holds
\beq
t^{(\l)}(x_p)t^{(\l)}(\om x_p)=\Phi_0(x_p;r_{1+\l,2+\l})&&+\nonumber\\
+\phi_1^-(x_p;r_{1+\l})^M&&
\ov t^{(\l)}(x_p)+
\phi_1^+(x_p;r_{1+\l})^M\ov t^{(\l)}(\om x_p),\quad\quad \label{f36}\\ 
\ov t^{(\l)}(x_p)\ov t^{(\l)}(\om x_p)=\Phi_0(x_p;r_{2+\l,1+\l})&&+
\nonumber\\
+\phi_1^-(x_p;r_{2+\l})^M&& t^{(\l)}(x_p)+
\phi_1^+(x_p;r_{2+\l})^M t^{(\l)}(\om x_p),\quad\quad\label{f37}
\eeq
with $\phi_1^{\pm}(x_p;r)$ defined in (\ref{f35}) and
\be
\Phi_0(x_p;r',r)=\l_1^M+\l_2^M+\l_3^M, \label{f38}
\ee
where
$\l_i$, $i=1,2,3$ are three roots of the following cubic equation
\be
\l^3+a\l^2+b\l+c=0, \label{f39}
\ee
\be
a=x_p^2\prod_{i=1}^3h_i^+(r)^2 \sum_{i=1}^3{h_i^+(r')^2\over h_i^+(r)^2}-
\prod_{i=1}^3h_i^-(r)^2 \sum_{i=1}^3{h_i^-(r')^2\over h_i^-(r)^2}, 
\label{f40}
\ee
\be
{\ds b=
\prod_{i=1}^3{h_i^+(r)^2\over [h_i^+(r')]^{-2}}
(\prod_{i=1}^3{h_i^-(r)^2\over h_i^+(r)^2}-x_p^2)}{\ds\Bigl
[\prod_{i=1}^3{h_i^-(r')^2\over h_i^+(r')^2}
\sum_{i=1}^3{h_i^-(r)^2\over h_i^-(r')^2}
-x_p^2\sum_{i=1}^3{h_i^+(r)^2\over h_i^+(r')^2}
\Bigr]},
\label{f41}
\ee
\be
c=(x_p^2\prod_{i=1}^3h_i^+(r')^2-\prod_{i=1}^3h_i^-(r')^2)
(x_p^2\prod_{i=1}^3h_i^+(r)^2-\prod_{i=1}^3h_i^-(r)^2))^2. \label{f42}
\ee

A system of functional equations similar (\ref{f36}-\ref{f37}) 
for ``factorized'' $sl(3)$ $L$-operators
has been obtained in \cite{K} and discussed in \cite{BB}. 
However, for that 
case there is no $Z_3$ symmetry with respect to
the indices of $L$-operators and as a result, the functional relations
will have a slightly more complicated form and involve explicitly
shift operators in the "quantum" space.

It is easy to see  that the system of functional equations 
(\ref{f36}-\ref{f37})
is invariant under the replacement $r_1\to r_2$, 
$t^{(\l)}(x_p;r_1,r_2)\to\ov t^{(\l)}(x_p;r_1,r_2)$.
However, it is not true in general that all eigenvalues of the
transfer matrices $t^{(\l)}(x_p)$, $\ov t^{(\l)}(x_p)$ satisfy 
$\ov t^{(\l)}(x_p;r_1,r_2)=t^{(\l)}(x_p;r_2,r_1)$.
The spectrum of $t^{(\l)}(x_p)$, $\ov t^{(\l)}(x_p)$ 
involves also ``non-symmetric'' solutions of (\ref{f36}-\ref{f37}).

\section{The homogeneous case $r_1=r_2$.}

A great simplification occurs if we restrict ourselves to the case
of the completely homogeneous model $r_1=r_2=r$. In this case all roots in 
(\ref{f39})
coincide and the function $\Phi_0(x_p;r,r)$ has a very simple form.
Let us define
\be
x_r=\prod_{i=1}^3{h_i^-(r)\over h_i^+(r)}. \label{f43}
\ee
It is easy to see that for this case the transfer matrices $t(x_p;r,r)$
and $\ov t(x_p;r,r)$ are effectively the functions of a single variable
$x=x_p/x_r$. Namely,
\be
{\ds t(x_p;r,r)=(x_r\prod_{i=1}^3 h_i^+(r))^Mt(x_p/x_r),}\quad 
{\ds \ov t(x_p;r,r)=(x_r\prod_{i=1}^3 h_i^+(r))^M\ov t(x_p/x_r).} 
\label{f44}
\ee
Then omitting the superscript $(\l)$ the system of functional equations 
(\ref{f36}-\ref{f37}) can be
rewritten as follows
\beq
&t(x)t(-x)=3(1-x^2)^M+(1-x)^M\ov t(x)+(1+x)^M \ov t(-x),&\nonumber\\
&\ov t(x)\ov t(-x)=3(1-x^2)^M+(1-x)^M t(x)+(1+x)^M t(-x),& \label{f45}
\eeq
where $t(x)$ and $\ov t(x)$ are polynomials in $x$ of the degree $M$.

Let us define
\be
u(x)=t(x)+(1+x)^M, \quad \ov u(x)=\ov t(x)+(1+x)^M. \label{f46}
\ee
Then we have from (\ref{f45}) 
\be
u(x)u(-x)=\ov u(x)\ov u(-x), \label{f47}
\ee
\be
u(x)u(-x)=(1-x)^M(u(x)+\ov u(x))+(1+x)^M(u(-x)+\ov u(-x)). \label{f48}
\ee

The solution of (\ref{f47}) can be written as follows
\be
u(x)=a(x)b(x),\quad \ov u(x)=a(-x)b(x),\quad
a(x)=\prod_{i=1}^k(u_i-x),\quad k=0,\ldots,M, \label{f50}
\ee
$u_i$, $i=1,\ldots,k$ - the roots of the polynomial $a(x)$ of the degree $k$
and $b(x)$ is the polynomial in $x$ of the degree $M-k$.

Then we have from (\ref{f47}-\ref{f48})
\be
c(x)c(-x)=(1-x^2)^M(a(x)+a(-x))^2, \label{f51}
\ee
\be
c(x)=a(x)a(-x)b(x)-(1+x)^M(a(x)+a(-x)), \label{f52}
\ee
where $c(x)$ is the polynomial in $x$ of the degree $M+k$.

In principle it is straightforward to solve (\ref{f51}-\ref{f52})
in terms of $u_i$. First we need to find the roots of the polynomial
$a(x)+a(-x)$:
\be
a(x)+a(-x)=\lambda\prod_{i=1}^{\lfloor k/2\rfloor}(v_i^2-x^2), \label{f53} 
\ee
where $\lfloor k/2\rfloor$ is the integer part of $k/2$. 

Now a general solution of (\ref{f51}) has the following form
\beq
&{\ds c(x)=\lambda(1-x)^n(1+x)^{M-n}\prod_{i=1}^{l_1}(v_i-x)
\prod_{i=l_1+1}^{\lfloor k/2\rfloor}(v_i+x)\times}\quad\quad& \nonumber \\
&{\times \ds\prod_{i=1}^{l_2}(v_i-x) 
\prod_{i=l_2+1}^{\lfloor k/2\rfloor}(v_i+x)},
\>\>\> n=0,\ldots,M,\>\> 0\leq l_1\leq l_2\leq \lfloor k/2\rfloor.&\label{f54}
\eeq
The last step is to substitute (\ref{f50}, \ref{f53}-\ref{f54})
into (\ref{f52}) and demand that $b(x)$ should be a polynomial in $x$.
It gives us a closed system of equations on $u_i$ and $v_i$.

As a result we come to the system of equations on $u_i$ and $v_i$:
\be
\prod_{j=1}^k{u_j-v_i \over u_j+v_i}=-1,\> i=1,\ldots,\lfloor k/2\rfloor,\quad
\prod_{j=1}^k(u_j+u_i)=\prod_{j=1}^{\lfloor k/2\rfloor}(v_j^2-u_i^2),\> i=1,\ldots,k,
\label{f56}
\ee
\be
\prod_{j=1}^{l_1}{v_j-u_i\over v_j+u_i}
\prod_{j=l_2+1}^{\lfloor k/2\rfloor}{v_j+u_i\over v_j-u_i}= -\Bigl({1+u_i\over 1-u_i}
\Bigr)^n,\>\>n=0,\>\>\ldots,M,l_1,l_2=0,\ldots,\lfloor k/2\rfloor. \label{f57}
\ee

It is not difficult to solve (\ref{f56}-\ref{f57}) for $k=0,1$ and to obtain
some eigenvalues of the transfer matrix $T_p\ov T_p$ at $r_1=r_2$
for any $M$. However, in general the system (\ref{f56}-\ref{f57})
is a transcendental system of equations and can not be solved explicitly
for any values of $k$. 

Therefore, (\ref{f50}-\ref{f57}) together with (\ref{f43}-\ref{f44}, \ref{f46})
give the solution for the spectrum of the transfer matrix $T_p\ov T_p$
in the homogeneous case $r_1=r_2$. However, we should say that 
this case is quite restrictive and, in fact, all dependence
on the vertical rapidity $r$ and matrices $K_{ij}$ in (\ref{l1})
can be absorbed by a redefinition of the parameter $x_p$.

\section{The Bethe ansatz technique for ``fused'' $L$-operators.}

In the previous section 
we considered the solution to the functional
relations (\ref{f36}-\ref{f37}) for the case 
$p_1=p_2=p,\> r_1=r_2=r$. However, a direct 
Bethe ansatz method can be also
developed for this case.
Let us make some transformations of the $L$-operators
given by (\ref{f15}-\ref{f17}) (here we consider only the case 
$\l = 0$).
Define the $L'$-operators:
\begin{equation}
L'_{ij} = \kappa\; \xi_i (p)/\xi_i (r)\;L^{(0)}_{ij}(p;r,r)\;
\xi_j (r)/\xi_j (p), \label{be1}
\end{equation}
where $\xi(p)$ is defined in (\ref{f26}), 
\be
\kappa = (h^+(p) h^-(p) h^+(r) h^-(r))^{-1/2},\quad
h^{\pm}(p) = \prod_{i=1}^3 h_i^{\pm}(p). \label{be3}
\ee
The transfer matrices for $L^{(0)}$- and $L'$-operators differ only by the 
scalar factor $\kappa^M$. Then all matrix elements
of $L'$ are the functions of a single parameter $x={(x_p/x_r)}^{1/2} $.
To obtain the  simplest form of $L'$ let us make the following 
equivalence transformation:
\begin{equation}
{\bar L}_{ij}\; =\; C\;L'_{ij}\;C^{-1}, \quad
C={1\over2}\;\;\left(\matrix{
1&{\;\;\;1}&{\;\;\;1}&{\;\;\;1}\cr
1&\;\;\;1&{-1}&-1\cr
1&{-1}&\;\;\;1&-1\cr
1&-1&-1&{\;\;\;1}\cr}\right)
\label{be5}
\ee
Then we have for the matrix elements of the $\bar L$-operators
\begin{eqnarray}
&{\bar L}_{ii} (\a,\b) & \;=\; \{x\} (\d_{\a,0}\d_{\b,0}-\d_{\a,i}\d_{\b,i}) +
[x] (\d_{\a,i+1} \d_{\b,i+1} - \d_{\a,i-1}\d_{\b,i-1}),\quad \label{Lii}\\
& {\bar L}_{ij} (\a,\b) & \;=\;2 x^{\e_{ij}} 
(\d_{\a,k}\d_{\b,0} - \d_{\a,j}\d_{\b,i}),
\label{Lij}
\end{eqnarray}
where $\{x\} = x+1/x, [x] = x-1/x$, the indices $i-1,i,i+1$ in (\ref{Lii}) 
are defined by the cyclic permutation of $1,2,3$ while 
$i,j,k$ in (\ref{Lij}) is any permutation of $1,2,3$ and 
\be
\e_{ij} = \cases{          1,     &$i<j;$\cr
                         -1,&$i>j.$\cr}      \label{e}
\ee
In  (\ref{Lii}-\ref{Lij}) we imply that the enumeration of rows
and columns is given by $(0,1,2,3)$.

From (\ref{Lii}-\ref{Lij}) we can easily observe that i)
the representation matrices of the diagonal $L$-operators are diagonal and 
for all of them the component $(0,0)$ is equal to $\{x\}$; ii)
 the non-diagonal $L$-operators have zeros in the $0$-rows.

Therefore, we can conclude that for the transfer matrix
\be
{\bar T}_{\a_1,\ldots,\b_M}^{\b_1,\ldots,\b_M} =\sum_{i_1,\ldots,i_M} 
{\bar L}_{i_1i_2}(\a_1,\b_1)\ldots {\bar L}_{i_Mi_1}(\a_M,\b_M) 
\label{transfer}
\ee
all matrix elements 
${\bar T}_{\a_1,\ldots,\a_M}^{j_1,\ldots,j_M} = 0$, for 
which there is at least one 
$0$-component among the set $\a_1,\ldots,\a_M$
and all indices $j_k$ run the  values $(1,2,3)$. So, the transfer matrix 
(\ref{transfer}) has the block-down-triangular form. 

For example, the simplest block has
a dimension $1$:
\be
{\bar T}_{0,\ldots,0}^{0,\ldots,0} = 3 \{x\}^M. \label{b0}
\ee

Further there are $M$ blocks $3\times 3$ of the form:
\be
{\bar T}_{0,\ldots,0,i_k,0,\ldots,0}^{0,\ldots,0,j_k,0,\ldots,0} = 
\{x\}^{M-1} {T^{(1)}}_{i_k}^{j_k}, \label{b1}
\ee
where $T^{(1)}$ is the one-site transfer matrix for the "reduced" $L$-operators
which can be obtained from ${\bar L}$ by removing all $0$-components, i.e.
\be
L_{ij} (i_1,j_1) = {\bar L}_{ij} (i_1,j_1)\quad i_1,j_1 = (1,2,3).  \label{LL}
\ee
The next $M (M-1)/2$ blocks in the transfer matrix have the 
dimension $3^2\times3^2$:
\be
{\bar T}_{0,\ldots,0,i_{k1},0,\ldots,i_{k2},0,\ldots,0}^
{0,\ldots,0,j_{k1},0,\ldots,j_{k2},0,\ldots,0} = 
\{x\}^{M-2} {T^{(2)}}_{i_{k1},i_{k2}}^{j_{k1},j_{k2}}. \label{b2}
\ee
where $T^{(2)}$ is the two-site transfer matrix for the "reduced" $L$-operators.
In the $n$-th step we have $C_M^n$ blocks obtained from
the $n$-site transfer matrices. In the last step we have one block:
${T^{(M)}}_{i_1,\ldots,i_M}^{j_1,\ldots,j_M}.$
As a result we obtain  a decomposition of the initial 
$4^M$-dimensional space into the direct sum
of the subspaces $\sum_{n=0}^M C_M^n 3^n$.
Therefore, the initial problem of calculation of the spectrum for 
the transfer matrix
(\ref{transfer}) has been reduced to the spectral problem for the 
transfer matrices 
$T^{(n)}$.

It is easy to see that the $L$-operators given by (\ref{LL}) coincide 
with the $R$-matrix (\ref{f29}) for the $sl(3)$ model.
Hence, one can use the standard nested Bethe ansatz technique \cite{V}
for this model. We give here only the final result for 
the eigenvalues for the deformed
$sl(3)$ model with the arbitrary deformation parameters $q$ and $\rho_{i,j}$.
\be
 \Lambda(x) = 
 {{[qx]}^n\over{\r_{1,3}^{b}\r_{2,3}^{a-b}}} \prod_{i=1}^a 
w({y_i\over x}) +
[x]^n \bigl[
{{\r_{2,3}^{n-a}}\over{\r_{1,2}^{b}}} {\prod_{i=1}^a w({x\over y_i}) }
{\prod_{k=1}^{b} w({z_k\over x})} + {\r_{1,3}^{n-a}\over \r_{1,2}^{b-a}} 
{\prod_{k=1}^{b} w({x\over z_k})}\bigr], \label{eig}
\ee
where $0\leq b\leq a\leq n$, $w(x)={{[q x]}\over{[x]}}$
and the two sets of parameters $y_1,\ldots,y_a$ and 
$z_1,\ldots,z_{b}$ should be defined from the system of the 
Bethe ansatz equations:
\beq
&w(y_i)^n = & (-1)^{a-1} \r_{2,3}^n \r^{-b}
\prod_{j\neq i}^a {{[q y_i/y_j]}\over{[q y_j/y_i]}}
{\prod_{k=1}^{b} w({z_k\over y_i})},\nonumber\\
&{\prod_{i=1}^a w({z_k\over y_i})} = 
&(-1)^{b-1}({{\r_{1,3}}\over{\r_{2,3}}})^n \r^a
{\prod_{l\neq k}^b {{[{q z_k/ z_l}]}\over{[{q z_l/ z_k}]}}}, 
\label{ba}
\eeq
where $\r =\r_{1,2}\r_{2,3}\r_{3,1}$.
However, we need only a  particular case of this solution  
\be
q=i, \quad\r_{j,j+1}=-\r_{j+1,j}=-i, \quad j=1,2,3.\label{be6}
\ee

In this case the formulas (\ref{ba}) become much more simple:
\be
w'(y_i)^n = (-1)^{n-a-1} \prod_{k=1}^{b} w'({z_k\over y_i}),\quad
\prod_{i=1}^a w'({z_k\over y_i}) = (-1)^{n-b-1} ,\quad
w'(x)={{\{x\}}\over{[x]}}. \label{ba1}
\ee
In comparison with the equations (\ref{f56}-\ref{f57}) 
the equations (\ref{ba1}) do not contain the redundant solutions. 
The reason for
this can be easily understood. Namely, if we consider another type of  
$L$-operators
which can be obtained from the initial ones by the transposition of the 
representation
matrices, one can conclude that the transfer matrix for 
them satisfies the same 
functional relations (\ref{f45}). 
The resulting equations of the Bethe ansatz for this case 
can be obtained from (\ref{ba})
by the following substitution:
\be
x\to x^{-1},y_i\to y_i^{-1},z_k\to z_k^{-1},\r_{i,j}\to\r_{j,i}.
\ee
After this we should fix the parameters of deformation as in (\ref{be6}).
So, the system of the equations (\ref{f56}-\ref{f57}) contains the solutions
to Bethe ansatz equations for both cases.
Unfortunetely, we have failed to find an explicit correspondence
between (\ref{f56}-\ref{f57}) and (\ref{ba1}).

\section{Discussion}

In this paper we have obtained the functional relations for
 eigenvalues of the transfer matrix of the $sl(3)$ chiral Potts model 
at $q^2=-1$. In the completely homogeneous case we
have also developed a direct Bethe ansatz scheme.

A whole set of functional equations for the usual chiral
Potts model has been obtained in \cite{BBP}. However, the $sl(3)$
case looks much more difficult. In particular, a proper
generalization of the Baxter's construction
of the $Q$-matrix for the eight-vertex model \cite{BaxB} which works for 
the usual chiral Potts model \cite{BS} fails for the $sl(3)$ case.
However, the structure of functional equations for the arbitrary
$N$ should be governed by a proper generalization of 
quadratic ``fusion'' rules \cite{KNS,KLWZ} which should have
the same functional form for the $sl(n)$ case at roots of unity as well.
Then it should be possible to define a set of $Q$-matrices
for the $sl(n)$ case. Of course, the boundary conditions 
for ``fusion'' rules and analytical properties of solutions 
at $q^N=-1$ will be extremely complicated.  
However, even a trigonometric limit of the $sl(n)$ chiral Potts
model at $q^N=-1$ is of a great interest, because it corresponds
to the $n$-layer Zamolodchikov model under the proper modification of
boundary conditions on a three-dimensional lattice.
We hope to address these problems in further publications.

\section{Acknowledgments}
The authors would like to thank F.C.~Alcaraz, V.~Bazhanov, M.~Batchelor,  
R.~Flume, G.~von~Gehlen, V.~Rittenberg,
Y.G.~Stroganov, P.~Wiegmann and 
Yu-Kui~Zhou 
for stimulating discussions and suggestions.
This research (VVM) has been  supported by the Australian Research Council
and (HEB) by Alexander~von~Humboldt Foundation. HEB would also like to
thank R.~Flume for his kind hospitality in the Physical Institute of
Bonn University.

\newpage
\begin{picture}(600,550)
\put(13,190){\Large\bf $T_{p_1}$}\put(13,265)
{\Large\bf $\ov T_{p_2}$}\put(50,80){
\begin{picture}(600,380)
\multiput(0,150)(150,0){3}{\usebox{\stars}}
\multiput(0,0)(150,0){3}{\usebox{\stars}}
\multiput(20,172.5)(0,150){2}{\usebox{\war}}
\multiput(570,97.5)(0,150){2}{\usebox{\ear}}
\multiput(176.5,10)(150,0){3}{\usebox{\sar}}
\multiput(101.5,410)(150,0){3}{\usebox{\nar}}
\multiput(11,112.5)(0,150){2}{\Large\bf $p_1$}
\multiput(585,187.5)(0,150){2}{\Large\bf $p_2$}
\multiput(112.5,5)(150,0){3}{\Large\bf $r_1$}\put(463,145){\large\bf$\s'_3$}
\multiput(187.5,435)(150,0){3}{\Large\bf $r_2$}
\put(65,45){\large\bf$\s_1$}\put(215,45){\large\bf$\s_2$}
\put(365,45){\large\bf$\s_3$}\put(515,45){\large\bf$\s_4$}
\put(163,145){\large\bf$\s'_1$}\put(313,145){\large\bf$\s'_2$}
\put(65,245){\large\bf$\s''_1$}\put(215,245){\large\bf$\s''_2$}
\put(365,245){\large\bf$\s''_3$}\put(515,245){\large\bf$\s''_4$}
\end{picture}}
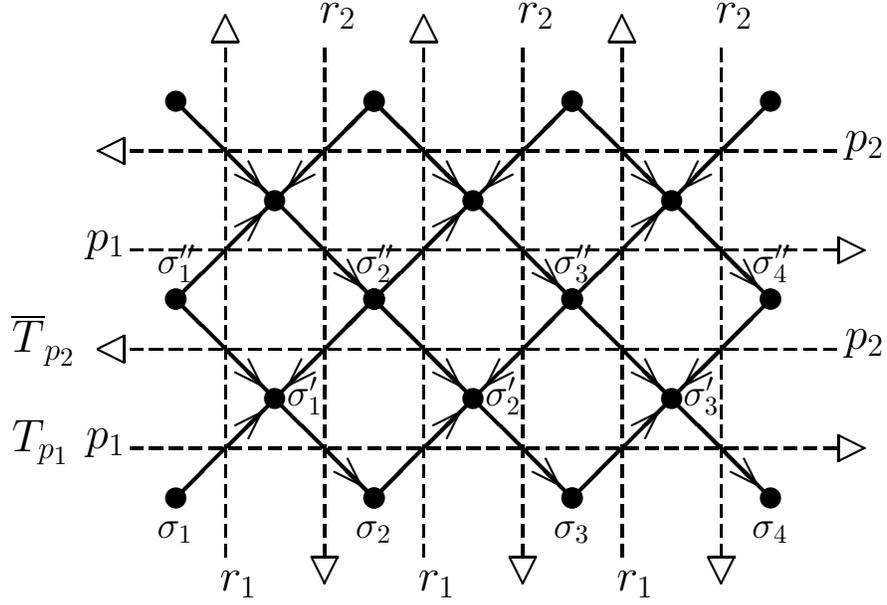
\put(110,20){\large\bf Figure 1. The square lattice ${\cal L}$.}
\end{picture}

\begin{picture}(600,250)
\put(25,70)
{\begin{picture}(600,150)
\thicklines
\multiput(0,0)(30,30){5}{\line(1,1){20}}
\multiput(150,0)(-30,30){5}{\line(-1,1){20}}
\put(120,120){\usebox{\near}}\put(0,120){\usebox{\nwar}}
\multiput(75,0)(0,150){2}{\circle*{15}}
\multiput(74,0)(1,0){3}{\line(0,1){150}}
\multiput(74,120)(1,0){3}{\line(-1,-2){10}}
\multiput(74,120)(1,0){3}{\line(1,-2){10}}
\put(0,30){\large $p$}\put(150,30){\large $q$}
\put(90,0){\large $\a$}\put(90,150){\large $\b$}
\put(140,75){\Large\bf {$=\> \overline W_{pq}(\a,\b)$}}
\end{picture}}
\put(355,70)
{\begin{picture}(600,150)
\thicklines
\multiput(0,150)(30,-30){5}{\line(1,-1){20}}
\multiput(150,150)(-30,-30){5}{\line(-1,-1){20}}
\put(0,0){\usebox{\swar}}\put(120,0){\usebox{\sear}}
\multiput(75,0)(0,150){2}{\circle*{15}}
\multiput(74,0)(1,0){3}{\line(0,1){150}}
\multiput(74,120)(1,0){3}{\line(-1,-2){10}}
\multiput(74,120)(1,0){3}{\line(1,-2){10}}
\put(0,30){\large $p$}\put(150,30){\large $q$}
\put(90,0){\large $\a$}\put(90,150){\large $\b$}
\put(140,75){\Large {$=\>{\displaystyle 1 \over 
{\displaystyle\phantom{I}{ \overline W_{qp}(\a,\b)}^{\phantom{I}}}}$}}
\end{picture}}
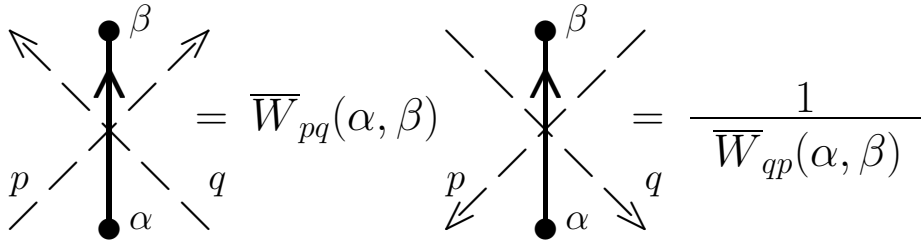
\put(50,15){\large\bf Figure 2. The weights $\ov W_{pq}(\a,\b)$ and
$\ov W_{qp}^{-1}(\a,\b)$}
\end{picture}

\begin{picture}(600,370)
\put(50,50){
\begin{picture}(600,300)
\multiput(30,112.5)(15,0){16}{\line(1,0){10}}
\multiput(30,187.5)(15,0){16}{\line(1,0){10}}
\multiput(112.5,30)(0,15){16}{\line(0,1){10}}
\multiput(187.5,30)(0,15){16}{\line(0,1){10}}
\multiput(75,150)(75,75){2}{\circle*{15}}
\multiput(150,75)(75,75){2}{\circle*{15}}
\multiput(74,150)(1,0){3}{\line(1,1){75}}
\multiput(74,150)(1,0){3}{\line(1,-1){75}}
\multiput(149,75)(1,0){3}{\line(1,1){75}}
\multiput(149,225)(1,0){3}{\line(1,-1){75}}
\multiput(97.5,10)(150,0){1}{\usebox{\sar}}
\multiput(172.5,255)(150,0){1}{\usebox{\nar}}
\put(5,172.5){\usebox{\war}}\put(255,97.5){\usebox{\ear}}
\put(0,112.5){\Large\bf $p_1$}\put(280,187.5){\Large\bf $p_2$}
\multiput(182.5,5)(150,0){1}{\Large\bf $r_1$}
\multiput(107.5,275)(150,0){1}{\Large\bf $r_2$}
\put(320,140){\Large\bf$=$}
\put(370,140){\Large\bf$\ov S_{\g,\a}^{\d,\b}(p_1,p_2;r_2,r_1)$}
\put(40,140){\Large\bf$\g$}\put(240,140){\Large\bf$\d$}
\put(140,245){\Large\bf$\b$}\put(140,45){\Large\bf$\a$}
\put(85,160){\usebox{\swar}}\put(185,160){\usebox{\sear}}
\put(110,85){\usebox{\sear}}\put(185,110){\usebox{\near}}
\end{picture}}
\put(100,15){\large\bf Figure 3. $R$-matrix $\ov S_{\g,\a}^{\d,\b}
(p_1,p_2;r_1,r_2)$}
\end{picture}

\begin{picture}(600,370)
\put(83,50){
\begin{picture}(600,300)
\multiput(0,0)(150,0){1}{\usebox{\stars}}
\put(20,172.5){\usebox{\war}}\put(270,97.5){\usebox{\ear}}
\multiput(176.5,10)(150,0){1}{\usebox{\sar}}
\multiput(101.5,260)(150,0){1}{\usebox{\nar}}
\put(10,112.5){\Large\bf $p_1$}\put(285,187.5){\Large\bf $p_2$}
\multiput(112.5,5)(150,0){1}{\Large\bf $r_1$}
\multiput(187.5,285)(150,0){1}{\Large\bf $r_2$}
\put(65,45){\large\bf$\a$}\put(227,45){\large\bf$\b$}
\put(167,145){\large\bf$\s$}
\put(65,245){\large\bf$\d$}\put(227,245){\large\bf$\g$}
\put(320,140){\Large\bf$=$}
\put(370,140){\Large\bf$W_{r_1r_2}^{p_1p_2}(\a,\b,\g,\d)$}
\end{picture}}
\put(100,15){\large\bf Figure
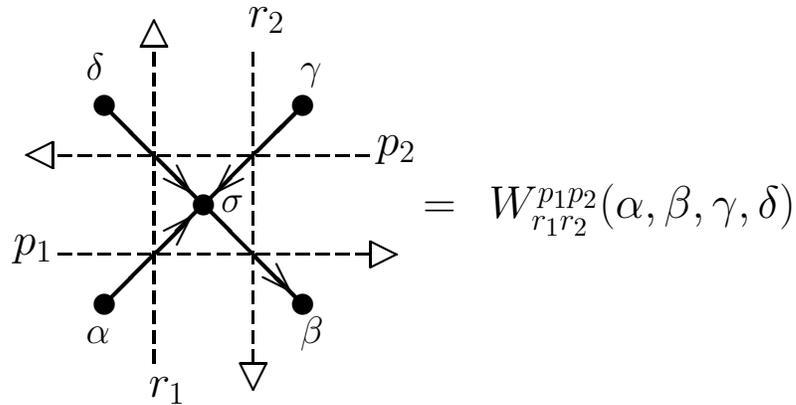
 4. The ``star'' weight $W_{r_1r_2}^{p_1p_2}(\a,\b,\g,\d)$}
\end{picture}

\begin{picture}(600,370)
\put(50,50){
\begin{picture}(600,300)
\multiput(0,0)(150,0){1}{\usebox{\starss}}
\put(20,97.5){\usebox{\war}}\put(270,172.5){\usebox{\ear}}
\multiput(101.5,10)(150,0){1}{\usebox{\sar}}
\multiput(176.5,260)(150,0){1}{\usebox{\nar}}
\put(10,187.5){\Large\bf $p_1$}\put(285,112.5){\Large\bf $p_2$}
\multiput(187.5,5)(150,0){1}{\Large\bf $r_1$}
\multiput(112.5,280)(150,0){1}{\Large\bf $r_2$}
\put(65,45){\large\bf$\a$}\put(227,45){\large\bf$\b$}
\put(167,145){\large\bf$\s$}
\put(65,245){\large\bf$\d$}\put(227,245){\large\bf$\g$}
\put(320,140){\Large\bf$=$}
\put(370,140){\Large\bf$\wid W_{r_2r_1}^{p_2p_1}(\a,\b,\g,\d)$}
\end{picture}}
\put(100,15){\large\bf Figure
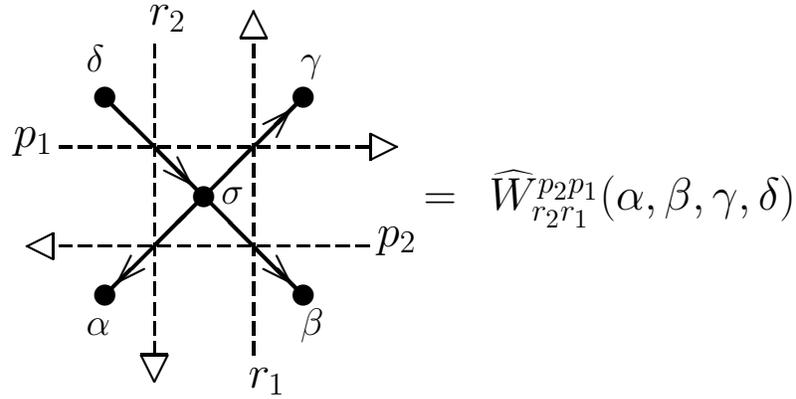
 5. The ``star'' weight $\wid W_{r_2r_1}^{p_2p_1}(\a,\b,\g,\d)$}
\end{picture}

\begin{picture}(600,400)
\put(25,50){
\begin{picture}(350,350)
\thicklines
\multiput(112.5,112.5)(15,0){11}{\line(1,0){10}}
\multiput(110,115)(-25,25){3}{\line(-1,1){20}}
\multiput(110,185)(-25,-25){3}{\line(-1,-1){20}}
\multiput(112.5,187.5)(15,0){11}{\line(1,0){10}}
\multiput(12.5,187.5)(15,0){2}{\line(1,0){10}}
\multiput(12.5,112.5)(15,0){2}{\line(1,0){10}}
\multiput(187.5,30)(0,15){11}{\line(0,1){10}}
\multiput(112.5,30)(0,15){11}{\line(0,1){10}}
\multiput(187.5,262)(0,15){2}{\line(0,1){10}}
\multiput(112.5,262)(0,15){2}{\line(0,1){10}}
\multiput(186.5,188)(-25,25){3}{\line(-1,1){20}}
\multiput(113.5,188)(25,25){3}{\line(1,1){20}}
\multiput(75,225)(150,0){2}{\circle*{15}}
\multiput(75,75)(150,0){2}{\circle*{15}}
\put(150,150){\circle*{15}}
\multiput(149,150)(1,0){3}{\line(1,1){75}}
\multiput(149,150)(1,0){3}{\line(1,-1){75}}
\multiput(74,75)(1,0){3}{\line(1,1){75}}
\multiput(74,225)(1,0){3}{\line(1,-1){75}}
\multiput(110,160)(75,-75){2}{\usebox{\sear}}
\put(110,110){\usebox{\near}}\put(160,160){\usebox{\swar}}
\multiput(74.66,75)(0.66,0){3}{\line(0,1){150}}
\multiput(75,224.66)(0,0.66){3}{\line(1,0){150}}
\multiput(65,105)(1,0){2}{\line(1,-2){10}}
\multiput(85,105)(1,0){2}{\line(-1,-2){10}}
\multiput(195,215)(1,0){2}{\line(2,1){20}}
\multiput(195,235)(1,0){2}{\line(2,-1){20}}
\put(65,45){\large\bf$\a$}\put(227,45){\large\bf$\b$}
\put(167,145){\large\bf$\s$}
\put(65,245){\large\bf$\d$}\put(227,245){\large\bf$\g$}
\put(-9,97.5){\usebox{\war}}\put(262,97.5){\usebox{\ear}}
\put(172.5,277){\usebox{\nar}}\put(172.5,10){\usebox{\sar}}
\put(-20,187.5){\Large\bf $p_1$}\put(280,187.5){\Large\bf $p_2$}
\multiput(112.5,5)(150,0){1}{\Large\bf $r_1$}
\multiput(112.5,297)(150,0){1}{\Large\bf $r_2$}
\end{picture}}
\put(400,50){
\begin{picture}(330,350)
\thicklines
\multiput(24.5,112.5)(15,0){11}{\line(1,0){10}}
\multiput(260,115)(-25,25){3}{\line(-1,1){20}}
\multiput(260,185)(-25,-25){3}{\line(-1,-1){20}}
\multiput(24.5,187.5)(15,0){11}{\line(1,0){10}}
\multiput(263,187.5)(15,0){2}{\line(1,0){10}}
\multiput(263,112.5)(15,0){2}{\line(1,0){10}}
\multiput(112.5,112.5)(0,15){11}{\line(0,1){10}}
\multiput(187.5,112.5)(0,15){11}{\line(0,1){10}}
\multiput(187.5,13)(0,15){2}{\line(0,1){10}}
\multiput(112.5,13)(0,15){2}{\line(0,1){10}}
\multiput(185.5,40)(-25,25){3}{\line(-1,1){20}}
\multiput(114.5,40)(25,25){3}{\line(1,1){20}}
\multiput(75,225)(150,0){2}{\circle*{15}}
\multiput(75,75)(150,0){2}{\circle*{15}}
\put(150,150){\circle*{15}}
\multiput(149,150)(1,0){3}{\line(1,1){75}}
\multiput(149,150)(1,0){3}{\line(1,-1){75}}
\multiput(74,75)(1,0){3}{\line(1,1){75}}
\multiput(74,225)(1,0){3}{\line(1,-1){75}}
\multiput(110,160)(75,-75){2}{\usebox{\sear}}
\put(185,185){\usebox{\near}}\put(85,85){\usebox{\swar}}
\multiput(224.66,75)(0.66,0){3}{\line(0,1){150}}
\multiput(75,74.66)(0,0.66){3}{\line(1,0){150}}
\multiput(215,105)(1,0){2}{\line(1,-2){10}}
\multiput(235,105)(1,0){2}{\line(-1,-2){10}}
\multiput(195,65)(1,0){2}{\line(2,1){20}}
\multiput(195,85)(1,0){2}{\line(2,-1){20}}
\put(3,97.5){\usebox{\war}}\put(277,97.5){\usebox{\ear}}
\put(172.5,262){\usebox{\nar}}\put(172.5,-8){\usebox{\sar}}
\put(65,45){\large\bf$\a$}\put(227,45){\large\bf$\b$}
\put(167,145){\large\bf$\s$}
\put(65,245){\large\bf$\d$}\put(227,245){\large\bf$\g$}
\put(-10,187.5){\Large\bf $p_1$}\put(295,187.5){\Large\bf $p_2$}
\multiput(112.5,-10)(150,0){1}{\Large\bf $r_1$}
\multiput(112.5,285)(150,0){1}{\Large\bf $r_2$}
\end{picture}}
\put(350,185){\Huge $=$}
\put(150,15){\large\bf Figure
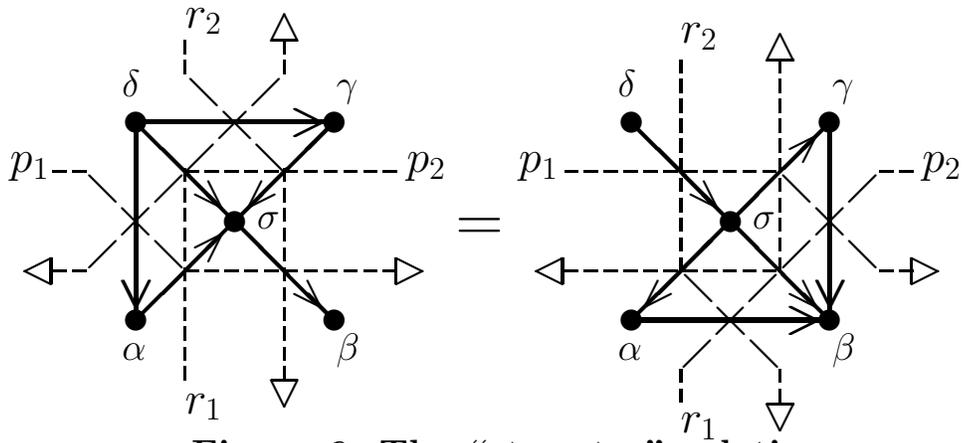
 6. The ``star-star'' relation.}
\end{picture}


\begin{thebibliography}{**}
\bibitem{YMP}
H. Au-Yang, B.M. McCoy, J.H.H. Perk, S. Tang and M.L. Yan,
{\it Phys. Lett.} {\bf 123A} (1987) 219-223.
\bibitem{MPTS}
B.M. McCoy, J.H.H. Perk, S. Tang and C.H. Sah, {\it Phys. Lett.}
{\bf 125A} (1987) 9-14.
\bibitem{BPY}
R.J. Baxter, J.H.H. Perk and H. Au-Yang, {\it Phys. Lett.} {\bf A128}
(1988) 138-142.
\bibitem{YP}
H. Au-Yang, J.H.H. Perk, {\it Proc. of Taniguchi Symposium}, v. {\bf 19}
(1989) Kinokumiya/Academic
\bibitem{BaxB}
R.J. Baxter, {\it ``Exactly Soluble Models in Statistical Mechanics''}
(1982) Academic, London.
\bibitem{BS}
V.V. Bazhanov, Yu. G. Stroganov, {\it Journ. Stat. Phys.}, {\bf 59}
(1990) 799.
\bibitem{L}
E.H. Lieb, {\it Phys. Rev.} {\bf 162} (1967) 162-172.
\bibitem{BKMS}
V.V. Bazhanov, R.M. Kashaev, V.V. Mangazeev and Yu.G. Stroganov,
Commun. Math. Phys. 138, 393-408 (1991).
\bibitem{DJMM}
E.Date, M. Jimbo, K. Miki and T. Miwa, {\it Commun. Math. Phys.}
{\bf 137} (1991) 133.
\bibitem{KMN}
R.M. Kashaev, V.V. Mangazeev, T. Nakanishi, {\it Nucl. Phys.} 
{\bf B362} (1991) 563
\bibitem{BB}
V.V. Bazhanov, R.J. Baxter, {\it Journ. Stat. Phys.}, v. {\bf 69}, 
n. 3/4, (1992) 453-485
\bibitem{Z1}
A.B. Zamolodchikov, {\it Zh. Eksp. Teor. Fiz.} {\bf 79} (1980) 641-664
[English transl.: {\it JETP} {\bf 52} (1980) 325-336]
\bibitem{Z2}
A.B. Zamolodchikov, {\it Commun. Math. Phys.} {\bf 79} (1981) 489-505.
\bibitem{BS1}
V.V. Bazhanov, Yu. G. Stroganov, {\it Teor. Mat. Fiz.} {\bf 52}
(1982) 105-113 [ English transl.: {\it Theor. Math. Phys.} {\bf 138}
(1982) 685-691]
\bibitem{JM}
M.T. Jaekel, J.M. Maillard, {\it Journ. Stat. Phys.} {\bf A15}
(1982) 1309.
\bibitem{BT}
R.J. Baxter, {\it Commun. Math. Phys.} {\bf 88} (1983) 185.
\bibitem{KMS}
R.M. Kashaev, V.V. Mangazeev and Yu.G. Stroganov, {\it 
Int. Journ. Mod. Phys.} {\bf A8}, (1993) 1399-1409.
\bibitem{Bax}
R.J. Baxter, {\it Physica} {\bf 18D} (1986) 321-347.
\bibitem{BQ}
R.J. Baxter, G.R.W. Quispel, {\it Journ. Stat. Phys.} {\bf 58}
n. 3/4, (1990) 411-430.
\bibitem{V}
P.P. Kulish, N.Yu. Reshetikhin, {\it Zh. Eksp. Teor. Fiz.}, {\bf 80},
no.1, (1981) 214.

O. Babelon, H.J. De Vega and C.M. Viallet, {\it Nucl. Phys} {\bf B220},
(1982) 266.
\bibitem{KRS}
P.P. Kulish, N.Yu. Reshetikhin and E.K. Sklyanin,
{\it Lett. in Math. Phys.} {\bf 5} (1981) 393-403.
\bibitem{KR}
A.N. Kirillov, N.Yu. Reshetikhin, {\it Journ. Phys. A: Math. Gen.}
{\bf 20} (1987) 1587.
\bibitem{BR}
V.V. Bazhanov, Yu.N. Reshetikhin, {\it Journ. Math. Phys. A: Math. Gen.}
{\bf23} (1990) 1477.
\bibitem{K}
R.M. Kashaev, Yu. G. Stroganov, private communication, 1991
\bibitem{CKP}
I.V. Cherednik, {\it Teor. Mat. Fiz}.{\bf 43} (1), 117-119 (1980);

P.P. Kulish, E.K. Sklyanin, {\it Zapiski Nauch. Semin. LOMI} {\bf95}, 129-160

Perk, J.H.H., Schultz, C.L., {Phys. Lett}.  A, {\bf 84}, 407 (1981)

\bibitem{BBP}
R.J. Baxter, V.V. Bazhanov and J.H.H. Perk, {\it Int. Journ. Mod. Phys.}
{\bf B4} (1990) 803-870.

\bibitem{KNS} 
A. Kuniba, T. Nakanishi and J. Suzuki, {\it Int. Journ. Mod. Phys.}
{\bf A9} (1994) 5215-5312.

\bibitem{KLWZ}I. Krichever, O. Lipan, P. Wiegmann, A. Zabrodin,
{\it Comm. Math. Phys.}, {\bf188} (1997) 267-304.

\end{thebibliography}
\end{document}